\documentclass[journal]{IEEEtran}
\usepackage{graphicx}
\usepackage{subfigure}
\usepackage{amsmath} 
\usepackage{amssymb} 
\usepackage{array}

\usepackage{blindtext}
\usepackage{color}

\newcommand{\mb}[1]{\mathbb{#1}}

\newcommand{\R}{\mb{R}}

\newcommand{\E}{\mb{E}}

\usepackage{mathrsfs}

\usepackage{amsmath}
\usepackage{algorithmic}
\usepackage{hyperref}

\def\BibTeX{{\rm B\kern-.05em{\sc i\kern-.025em b}\kern-.08em
    T\kern-.1667em\lower.7ex\hbox{E}\kern-.125emX}}

\usepackage{mwe}
\usepackage{fancyhdr}

\begin{document}
%
\title{Modeling satellite-based open water fraction via flexible Beta regression:
An application to wetlands in the north-western Pacific coast of Mexico}

\author{\IEEEauthorblockN{Inder~Tecuapetla-G\'omez,
Julia Trinidad Reyes}
\thanks{
I.~Tecuapetla-G\'omez,
Direcci\'on de C\'atedras,
Consejo Nacional de Ciencia y Tecnolog\'ia (CONACyT),
Av.~Insurgentes Sur 1582, Cr\'edito Constructor, Benito Ju\'arez 03940,
Ciudad de M\'exico.

I.~Tecuapetla-G\'omez,
Direcci\'on de Geom\'atica,
Comisi\'on Nacional para el Conocimiento y Uso de la Biodiversidad
(CONABIO), Liga Perif\'erico-Insurgentes Sur 4903,
Parques del Pedregal, Tlalpan 14010, Ciudad de M\'exico (e-mail: itecuapetla@conabio.gob.mx).

J. Trinidad Reyes,
Licenciatura en Matem\'aticas Aplicadas,
Facultad de Ciencias, UNAM,
Ciudad de M\'exico (e-mail:july\_treyes@ciencias.unam.mx).
}}

\maketitle

\renewcommand{\headrulewidth}{0in}
\pagestyle{empty}

\pagestyle{fancy}
\pagenumbering{gobble}

\begin{abstract}
Carbon sequestration and water filtering are two examples of the several
ecosystem services provided by wetlands. 
Open water mapping is an effective means to measure any wetland
extension as these are comprised of many open water bodies.
An economical, though indirect, approach towards mapping open water bodies 
is through applying geo-computational methods to satellite images.
In this work we propose the flexible Beta regression (FBR) model
to predict open water fraction from measurements of a water index. 
We focus on observations derived from two MODIS
images acquired during the dry season of 2008 in
Marismas Nacionales, a wetland located in the north-western
Pacific coast of Mexico. 
A Bayesian estimation procedure is presented to estimate
the FBR model; in particular, we provide details of a
nested Metropolis-Hastings and Gibbs sampling algorithm to carry
out parameter estimation. Our results show that the FBR model
produces valid predictors of water fraction unlike the standard
model. Our work
is complemented by software developed in the R language and
available through a GitHub repository.
\end{abstract}


%
\IEEEpeerreviewmaketitle

\section{Introduction and Data}\label{sec.intro}

Wetlands are ecosystems which serve as feeding and refuge areas for fish and
crustaceans. Also, they function as biological water filters, act as natural 
systems to control floods, provide a natural environment for the sequestration
and long-term storage of carbon dioxide from the atmosphere; they are
the largest natural source of methane to the atmosphere 
\cite{mitsch2013, kirschke2013}.  

In light of the above, modeling the dynamic of these ecosystems has gained a lot
of attention in recent years. An indirect and economical way to study the dynamic 
of extensive wetlands is through modeling biophysical variables derived from
satellite images \cite{prigent2001}. 
For instance, in order to determine the seasonal dependent
extension of a wetland, time series of moisture or water indices can be
generated and analyzed for the detection of open water. 
Typically, these indices are obtained by combining different spectral bands of satellite products such as the Moderate Resolution Imaging Spectroradiometer 
(MODIS) or the Landsat mission.

Since the spatial resolution of a Landsat image (30 m) is higher than
that of many MODIS products (250, 500 or 1000 m), it is sensible to classify
open water in a Landsat image and then calculate the corresponding water
fraction in the MODIS image. For instance, if a MODIS pixel contains 25 Landsat
pixels and 12 of the latter are classified as open water, then the fraction
of water in the MODIS pixel is 12/25. The success of this seemingly straightforward procedure depends on acquiring Landsat images with good quality data 
(e.g., cloudless and free of shadow clouds and artifacts). It is common 
that during an entire dry season, only one Landsat image meets this
quality requirements.
This precludes the
generation of time series of water fraction of MODIS images from previously
classified Landsat products. Generating MODIS time series of water indices, 
however, is a simple task.
In \cite{colditz2018} the authors found that among 14 moisture and water indices, 
the MODIS-based Modified Normalized Difference Water Index $\mbox{B}_6$ (MNDWI6), 
see \cite{xu2006}, has the best performance in appropriately defining open 
water bodies based on moderate spatial resolution products (such as Landsat).

\begin{figure}
  \centering  
  \subfigure{\includegraphics[height=4.5cm,width=3cm]{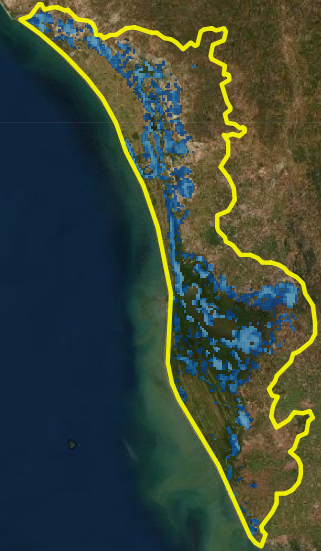}
  }
  \qquad
  \subfigure{\includegraphics[height=4.5cm,width=3cm]{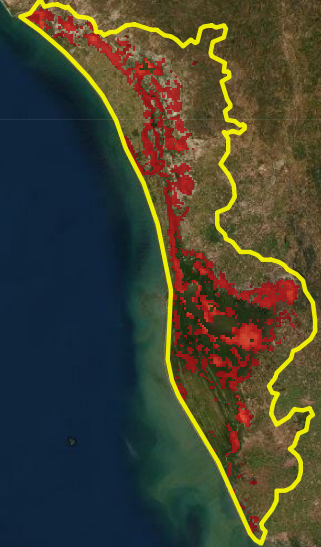}
  }
  \caption{Marismas Nacionales (in yellow).
  (Left) Pixels of water fraction with good quality assessment (in blue). (Right) Pixels of MNDWI6 with good quality assessment (in red).}\label{fig_maps}
\end{figure}

It is of interest to assess whether the MNDWI6 is an appropriate predictor of 
MODIS-based open water fraction 
which, in turn, was derived from a reference Landsat image.
Should the MNDWI6 be a suitable predictor
of open water then we shall use time series of this product for open water
mapping thus circumventing the acquisition of spotless Landsat images. 

In this work we provide an approach for modeling water fraction
as a function of MNDWI6. Our application is tested on
2270 observations with good quality assessment.
These observations were taken from 
two MODIS images with spatial resolution of 500 m, see Figure~\ref{fig_maps}. 
These images were acquired on the 82 day of 2008, during
the dry season of that year. The water fraction image was derived from
a Landsat one (acquired on the same date) which was used to classify
water presence; expert classification was applied to the Landsat image.
These images include Marismas Nacionales whose
133,854 ha of estuaries, small patches of 
cedars, oaks, amapas and oil palm, coconut palm and white, red,
black and Chinese mangrove, among other vegetation types,
makes it the most important wetland in north-western Mexico; 
Marismas Nacionales is located at
$22^\circ07\mbox{'}N$ and $105^\circ31\mbox{'}W$ \cite{conabio2016}.

\begin{figure}[htbp]
\centerline{\includegraphics[width=0.5\textwidth]{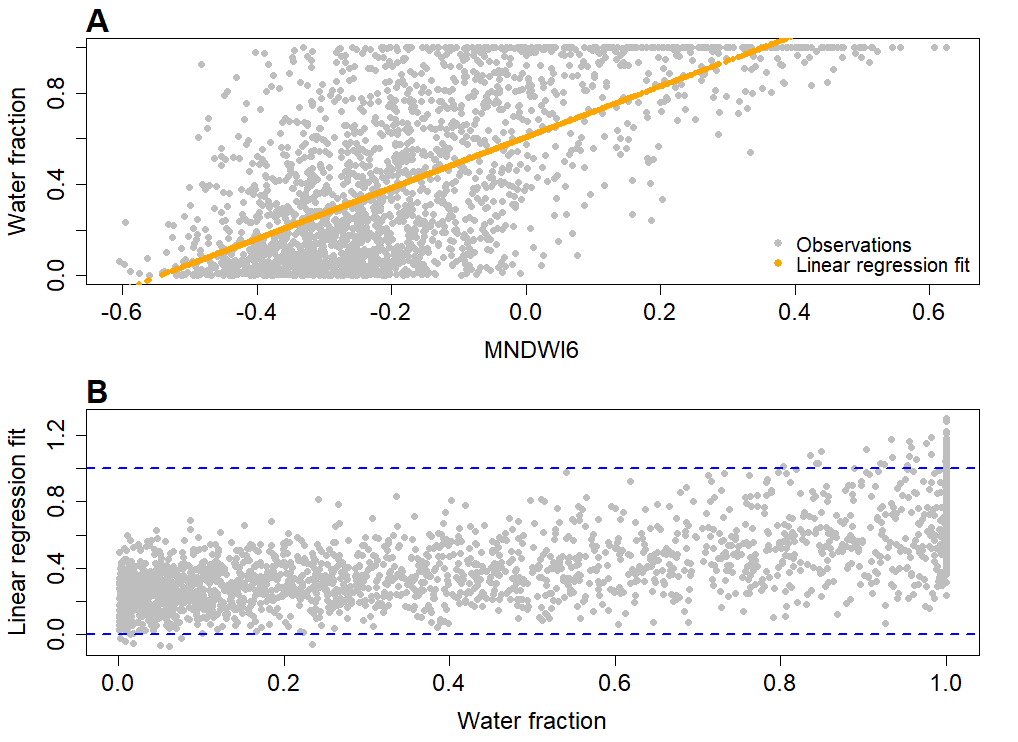}}
\caption{\textbf{A} Linear regression of MNDWI6 and water fraction. 
\textbf{B} Scatterplot of water fraction vs. fitted values.}
\label{fig_linearModel}
\end{figure}

A first modeling approach consists of fitting a linear regression between 
the water fraction and the MNDWI6.
This model yields the line displayed in 
Figure~\ref{fig_linearModel}-\textbf{A}; 
the intercept and slope estimates of this model are 
$\hat{\beta}_0^{\mbox{\tiny{lrm}}}=0.6041$ and
$\hat{\beta}_1^{\mbox{\tiny{lrm}}}=1.1142$, respectively. 
A scatterplot of the fitted values of this model against the water fraction 
variable, see Figure~\ref{fig_linearModel}-\textbf{B}, reveals that this approach 
is not appropriate for modeling purposes; some of the values predicted
by the linear model lie outside the interval $(0,1)$.

In what follows we describe the novel flexible Beta regression (FBR) model which, by construction, takes into account that  the response variable (water fraction) is bounded, hence, improving upon the linear model.
Section II presents the basics of the FBR model,
as well as details of the Bayesian estimation procedure employed to fit the FBR;
a nested Metropolis-Hastings and Gibbs sampling algorithm is discussed. 
The routines needed for this algorithm were developed in the R language \cite{RCore}
and are available at the Github repository \href{https://github.com/inder-tg/FBR}{https://github.com/inder-tg/FBR}.
Results are presented in Section III and we provide an outlook of future
work in the concluding Section IV.

\section{Methodology}\label{sec.methods}

Let $\boldsymbol{Y}^\top=(Y_1,\ldots,Y_n)$ denote $n$ independent response variables, each one
taking values on the interval $(0,1)$. 
Flexible Beta regression was introduced by \cite{migliorati2018} as an alternative for
modeling responses like $\boldsymbol{Y}$, which are bounded, as a function of some
covariates. Briefly, we will discuss the main features of the flexible Beta regression model.

\subsection{On flexible Beta random variables}

The density function of a flexible Beta random variable (rv) is given by
\begin{equation}\label{eq_fBeta_density}
	f_{FB}^\ast( y; \lambda_1, \lambda_2, \phi, p)
	=
	p\,f_B^\ast(y;\lambda_1,\phi) + (1-p)\,f_{B}^{\ast}(y;\lambda_2,\phi),
\end{equation}
where $y\in (0,1)$, $0<\lambda_2<\lambda_1<1$, $\phi>0$, $p\in (0,1)$ 
and $f_B^\ast$ denotes the density of a mean-precision parametrized Beta rv:
\[
	f_B^\ast(x;a,b)
	=
	\frac{\Gamma(b)}{\Gamma(ab)\,\Gamma((1-a)b)}\,x^{ab-1}(1-x)^{(1-a)b-1},
\]
where $x\in(0,1)$, $a\in (0,1)$, $b>0$.

That is, $f_{FB}^{\ast}$ is a mixture of two mean-precision Beta rvs with different
means ($\lambda_1$ and $\lambda_2$) but with the same precision parameter ($\phi$).
It is not difficult to see that the expected value of a rv with density function given 
by Eq.\eqref{eq_fBeta_density} is equal to $p\lambda_1 + (1-p)\lambda_2$. The latter,
gives rise to a reparametrization of the flexible Beta: $\mu=p\lambda_1 + (1-p)\lambda_2$,
$\phi=\phi$, $\tilde{\omega}=\lambda_1-\lambda_2$ and $p=p$. This reparametrization
has proven useful in enhancing parameter interpretability in the flexible Beta regression model.

The flexible Beta density has a wide range of shapes, from unimodal, monotone and U-shaped
to heavy-tailed. This characteristic makes it attractive to describe the main features
of a broad group of variables with applications in several scientific fields. In addition
to this, under mild conditions, it is known that two elements of the FB parametric
family are equal if and only if the corresponding parameters are the same (see Proposition~1
of \cite{migliorati2018}), this type of strong identifiability property is typically
found in non-mixture models.

\subsection{The flexible Beta regression model}

Let us suppose that the $i$-th response variable, $Y_i\in (0,1)$, is linked to the vector 
of covariates $\boldsymbol{x}_i\in \R^{k}$. The flexible Beta regression (FBR) model consists of 
assuming that each $Y_i$ is independently distributed flexible Beta; in notation, 
$Y_i \sim FB(\mu_i,\phi,\omega,p)$ where for some \emph{link} function $g:[0,1]\to \R$,
\begin{equation}\label{eq_FBR_model}
	g(\mu_i) = \E [ Y_i \mid \boldsymbol{x}_i ] = \boldsymbol{x}_i^\top\,\boldsymbol{\beta},
	\quad
	\boldsymbol{\beta}\in\R^k,
\end{equation}
and $\tilde{\omega} = \omega\,\min\{ \mu_i/p, (1-\mu_i)/(1-p)\}$. Observe that this
representation allows the parameters $\mu_i$, $\omega$ and $p$ to vary freely on $(0,1)$ 
and $\phi>0$. The parameter $\boldsymbol{\beta}$ is unknown. 

Note also that the representation induced by the FBR model is equivalent 
to requiring that
\begin{align}\label{eq_lambda1_lambda2}
	\lambda_1 
	&= 
	\mu_i + (1-p)\,\tilde{\omega} \notag\\
	\lambda_2
	&=
	\mu_i - p\,\tilde{\omega},
\end{align}
in Eq.\eqref{eq_fBeta_density}. So that, in what follows we will use the notation
$f_{FB}^\ast(\cdot;\mu_i,\phi,\omega,p)$ to refer to the density associated to 
an observation of the FBR model. For our application, we will consider the \emph{logit}
as the link function in Eq.\eqref{eq_FBR_model}. We remark that the flexible Beta regression
is not a generalized linear model as the distribution of mixture
of Betas does not belong to the exponential family.

\subsection{Bayesian estimation}

Let $\boldsymbol{y}^\top=(y_1,\ldots,y_n)$ denote $n$ independent observations of the FBR 
model just introduced. The likelihood function of this model is by definition
\begin{equation}\label{eq_likelihood}
	\mathscr{L}( \boldsymbol{\eta} \mid \boldsymbol{y} )
	=
    \prod_{i=1}^n\,f_{FB}^\ast(y_i\mid \mu_i, \phi, \omega, p),
\end{equation}
where $\boldsymbol{\eta}=(\boldsymbol{\beta},\phi,\omega,p)$. According to Proposition~2 of 
\cite{migliorati2018}, this function is bounded from above almost surely, which ensures the 
existence of a finite global maximum on the parameter space. This result provides
computational tractability to the FBR model, in particular for Bayesian estimation algorithms.

The underlying assumption of the FBR is that there are two groups of observations and the mean
of one of these groups dominates the other, see Eq.\eqref{eq_lambda1_lambda2}. Given the
$i$-th observation, however, we do not know to which of the two mixture components belongs. 
This makes the optimization of Eq.\eqref{eq_likelihood} unfeasible. To ameliorate 
this, a Bayesian approach is chosen, in particular, a data augmentation one is devised.
More precisely, define the $n$-dimensional random vector $\boldsymbol{v}$ with entries
\[
	v_i
	=
	\begin{cases}
		1 & \mbox{ if } y_i \mbox{ belongs to the first mixture component,}\\
		0 & \mbox{ otherwise }
	\end{cases}.
\]
Although essentially these latent variables ($\boldsymbol{v}$) are missing data,
they can be included in the model and used for parameter estimation in a 2-step
algorithm. In the first step, the parameter posterior distribution is obtained
conditional on $\boldsymbol{v}$, and in the second step, the observations are 
classified ($\boldsymbol{v}$ is updated) conditional on knowing the parameter. 

A key element in this algorithm is the \emph{complete-data} likelihood function:
{\footnotesize
\begin{align}
	\mathscr{L}_{CD}(\boldsymbol{\eta} \mid \boldsymbol{y}, \boldsymbol{v}) 
	= 
	\prod_{i=1}^n\,\left[ p f_{B}^\ast( y_i; \lambda_{1}, \phi ) \right]^{v_i}
	\left[(1-p) f_B^\ast(y_i; \lambda_{2}, \phi) \right]^{(1-v_i)},
\end{align}}%
where $\lambda_1$ and $\lambda_2$ are given by \eqref{eq_lambda1_lambda2}. Thus,
given an appropriate \emph{prior} distribution $\pi(\boldsymbol{\eta})$, the resulting
\emph{posterior} distribution is:
\[
	\pi(\boldsymbol{\eta},\boldsymbol{v}\mid \boldsymbol{y})
	\propto
	\mathscr{L}_{CD}(\boldsymbol{\eta} \mid \boldsymbol{y}, \boldsymbol{v}) \, \pi(\boldsymbol{\eta}).
\]

Since there is no prior information available about the response variables utilized in our
application, and, as we argued above, the parameters of the FBR model are strongly
identifiable (also known as variation independent), it is sensible to assume a factorized 
joint prior distribution:
\begin{equation}\label{eq_prior}
	\pi(\boldsymbol{\eta})
	=
	\pi(\boldsymbol{\beta})\,\pi(\phi)\,\pi(\omega)\,\pi(p).
\end{equation}
Following \cite{migliorati2018} and \cite{bouguila2006}, in our application, 
we used a $k$-dimensional Gaussian
prior $\boldsymbol{\beta}\sim\mathscr{N}_k(\boldsymbol{0},\Sigma_{\boldsymbol{\beta}})$ for the regression
parameters; we employed a gamma prior distribution $\phi \sim \mathscr{G}(g,g)$ for
the precision parameter, a rather usual choice, see e.g. \cite{branscum2007}; and we chose 
non-informative uniform priors for the parameters $\omega \sim \mathscr{U}(0,1)$ and $p \sim \mathscr{U}(0,1)$.
The $k\times k$ matrix $\Sigma_{\boldsymbol{\beta}}$ has \emph{large} values in its main diagonal and zeroes
otherwise. In the next section we provide details about the algorithm that allows us
to fit the FBR model based on the complete-data approach just presented.

\subsection{Nested Metropolis-Hastings and Gibbs sampling}\label{ss_MH_Gibbs}

Some parameters of the FBR model have known conditional posterior distributions, whereas the posterior
distribution of others have to be calculated numerically. For instance, it is not difficult 
to see that
\begin{equation}\label{eq_posterior_v}
	\pi( v_i \mid y_i, \boldsymbol{\eta} ) 
	= 
	\mbox{Bernoulli}( \pi( v_i=1 \mid y_i, \boldsymbol{\eta} ) ),
\end{equation}
where
\[
	\pi( v_i=1 \mid y_i, \boldsymbol{\eta} )
	=
	\frac{p\,f_B^\ast( y_i\mid \lambda_1,\phi )}{p\,f_B^\ast( y_i\mid \lambda_1,\phi ) + (1-p)\,f_B^\ast( y_i\mid \lambda_2,\phi )}.
\]
Note that by setting $n_0= \#\{t \mid v_t=0\}$ and $n_1= \#\{s \mid v_s=1\}$,
\begin{equation}\label{eq_posterior_p}
	\pi(p \mid \boldsymbol{v})
	=
	\mbox{Beta}(n_1+1,n_0+1).
\end{equation}
Also, we get 
\begin{align}\label{eq_posterior_omega}
	\pi( \tilde{\omega} \mid \boldsymbol{\beta}, p )
	&=
	\mathscr{U}(0,r(\boldsymbol{\beta},p)),\notag \\
	r(\boldsymbol{\beta},p)
	&=
	\min\left\{ \frac{\mu_i(\boldsymbol{\beta})}{p}, \frac{1-\mu_i(\boldsymbol{\beta})}{1-p}\right\},\notag \\
	\mu_i(\boldsymbol{\beta})
	&=
	\frac{\mbox{exp}( \boldsymbol{x}_i^\top\,\boldsymbol{\beta} )}{1+\mbox{exp}( \boldsymbol{x}_i^\top\,\boldsymbol{\beta} )}
	:=
	\mbox{logit}^{-1}(\boldsymbol{x}_i^\top\,\boldsymbol{\beta}),\notag \\
	\mbox{logit}(x)
	&=
	\log\left(x/(1-x)\right),\quad x\in(0,1).	
\end{align} 

Samples from the conditional posterior distribution of $(\phi, \boldsymbol{\beta})$ given 
$(\boldsymbol{y},\boldsymbol{v},p,\tilde{\omega})$ are obtained through Metropolis-Hastings (M-H)
algorithms, see Ch.~10 of \cite{gelman2013}. In particular, for the conditional posterior
of $\phi$ we employed a random walk as the states transfer (jumping) distribution. For the conditional
posterior of $\boldsymbol{\beta}$ the jumping distribution is a bivariate normal with mean
equal to the value of the chain in the previous step and covariance matrix $\Sigma_J$,
a diagonal matrix. 

The following pseudocode summarizes the main steps needed to obtain
samples from the conditional posterior distribution $\pi(\boldsymbol{\eta},\boldsymbol{v}\mid \boldsymbol{y})$.

\begin{algorithmic}
	\REQUIRE $\boldsymbol{x}$, $\boldsymbol{y}$, $\mbox{nSamples}$, $p_0$, 
	$\omega_0$, $\phi_0$, $\boldsymbol{\beta}_0$, 
	$\sigma_\phi$, $g$, $\Sigma_J$, $\Sigma_{\boldsymbol{\beta}}$
	\ENSURE $\boldsymbol{\mu}^{(0)}=(\mbox{logit}^{-1}(\boldsymbol{x}_1^\top\,\boldsymbol{\beta}_0),\ldots,\mbox{logit}^{-1}(\boldsymbol{x}_n^\top\,\boldsymbol{\beta}_0))$,\\
	$\boldsymbol{\omega}^{(0)} = \omega_0 \cdot \min\{ \boldsymbol{\mu}^{(0)}/p_0, (1-\boldsymbol{\mu}^{(0)})/(1-p_0)\}$
	\FOR{$j=1$ \TO $\mbox{nSamples}$}
	\STATE $v_{\mbox{\tiny{posterior}}} \leftarrow \pi( \boldsymbol{v} \mid \boldsymbol{y}, 
	\boldsymbol{\omega}^{(j-1)}, p_{j-1}, \boldsymbol{\mu}^{(j-1)},\phi_{j-1} )$
	\STATE $p_j \leftarrow \pi(p \mid v_{\mbox{\tiny{posterior}}})$
	\STATE $\boldsymbol{\mu}^{(j)} \leftarrow (\mbox{logit}^{-1}(\boldsymbol{x}_1^\top\,\boldsymbol{\beta}_{j-1}),\ldots,\mbox{logit}^{-1}(\boldsymbol{x}_n^\top\,\boldsymbol{\beta}_{j-1}))$
	\STATE $\boldsymbol{\omega}^{(j)} \leftarrow \pi(\tilde{\omega}\mid \boldsymbol{\mu}^{(j)})$
	\STATE $\phi_i \leftarrow \pi_{\mbox{\tiny{M-H}}}( \phi \mid \boldsymbol{y},v_{\mbox{\tiny{posterior}}},
	\boldsymbol{\mu}^{(j)},\boldsymbol{\omega}^{(j)},p_j,\phi_{j-1},\sigma_{\phi},g)$
	\STATE $\boldsymbol{\beta}_{j} \leftarrow \pi_{\mbox{\tiny{M-H}}}( \boldsymbol{\beta} \mid \boldsymbol{x}, 
	\boldsymbol{y}, v_{\mbox{\tiny{posterior}}}, p_j, \omega_i, \phi_j, \boldsymbol{\mu}^{(j)},
	 \boldsymbol{\beta}_{j-1}, \Sigma_J, \Sigma_{\boldsymbol{\beta}} ) $ 
	\ENDFOR
\end{algorithmic}
Details about burning periods, thinning, chain size and values of initial parameters
of this algorithm will be provided in the next section.

Having the sequence of vectors $\boldsymbol{\beta}_{j}$, 
$j=1,\ldots,\mbox{nSamples}$, which are
appropriate samples of the posterior distribution of the regression parameter 
$\boldsymbol{\beta}$,
we computed the median of each column to estimate the coefficients of the FBR model \eqref{eq_FBR_model}:
\begin{align*}
	\hat{\beta}_r
	&= 
	\mbox{median}(\boldsymbol{\beta}_{j}[,r]),\quad r=1,\ldots,k.
\end{align*}
Similarly, $\hat{p}=\mbox{median}(p_1,\ldots,p_{\mbox{\tiny{nSamples}}})$,
$\hat{\boldsymbol{\mu}}=(\mbox{logit}^{-1}(\boldsymbol{x}_1^\top\,\hat{\boldsymbol{\beta}}),\ldots,\mbox{logit}^{-1}(\boldsymbol{x}_n^\top\,\hat{\boldsymbol{\beta}}))$
and $\tilde{\omega}_{\mbox{\tiny{opt}}}=\min\{ \hat{\boldsymbol{\mu}}/\hat{p}, (1-\hat{\boldsymbol{\mu}})/(1-\hat{p})\}$.
Consequently, substituing these estimators in \eqref{eq_lambda1_lambda2}
we obtained the regression functions $\hat{\lambda}_1$ and
$\hat{\lambda}_2$.



\section{Results}\label{sec.results}


For our application we considered $n=2270$ observations of the water fraction product 
as the response variable ($y_i$) and an equal number of values of the MNDWI6 index 
as their corresponding covariates ($z_i$); each $y_i$ and $z_i$ were recorded at 
the same geographical position (pixel). We included an intercept in our regression model, hence, $\boldsymbol{x}_i^\top=(1,z_i)$, that is, in the notation above, $k=2$. 



For sampling from the conditional posterior distributions, we ran initially  
20 chains each having size $\mbox{nSamples}=8000$. 
We used $\boldsymbol{\beta}_0=(\hat{\beta}_0^{\mbox{\tiny{lrm}}},\hat{\beta}_1^{\mbox{\tiny{lrm}}})$. 
Our results were stable across different values of $p_0$ and $\omega_0$;
we used small values for the diagonal of $\Sigma_J$ whereas large values
for the diagonal of $\Sigma_{\boldsymbol{\beta}}$. 
We tried different starting values for the chain of the precision parameter
$\phi$. We found stable chains when $\phi_0$ was in a vicinity of 3, 
$\sigma_\phi=0.125$ and instead of using a $\mbox{Gamma}(g,g)$ we employed a 
$\mbox{Gamma}(\kappa\,g,g)$ where $\kappa=30,35$ and $g=0.1,0.2,0.3$.

\begin{figure}[htbp]
\centerline{\includegraphics[width=0.5\textwidth]{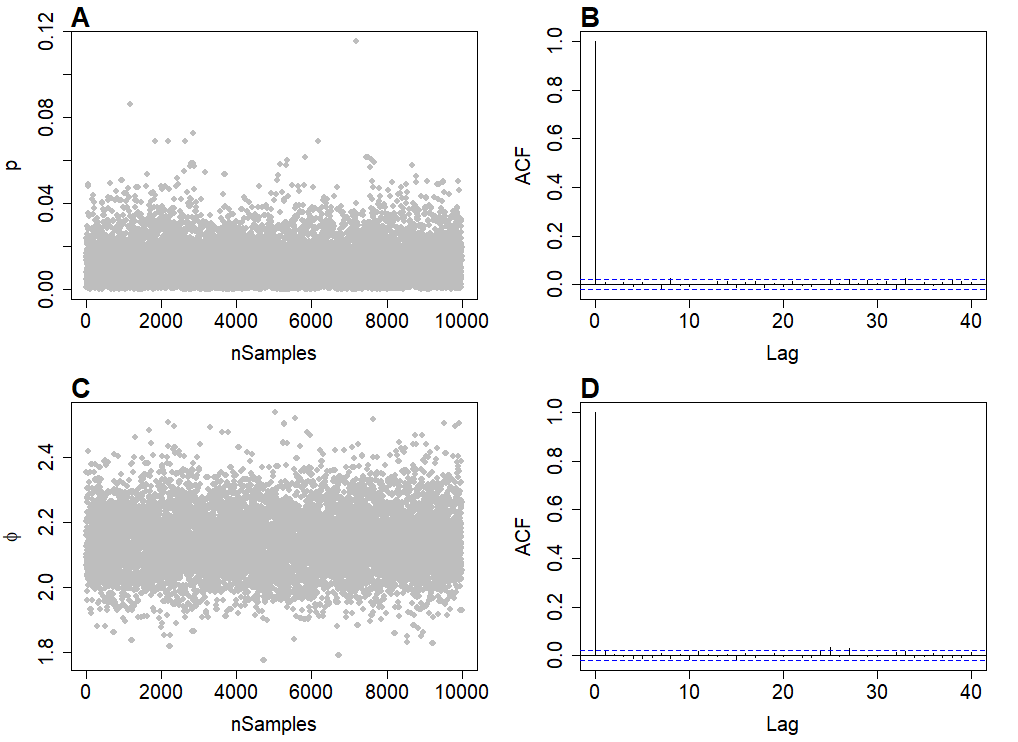}}
\caption{\textbf{A}-\textbf{B} Posterior samples of parameter $p$ with
ACF. \textbf{C}-\textbf{D} Posterior samples of precision parameter $\phi$
with ACF.}
\label{fig_p_phi}
\end{figure}

\begin{figure}[htbp]
\centerline{\includegraphics[width=0.5\textwidth]{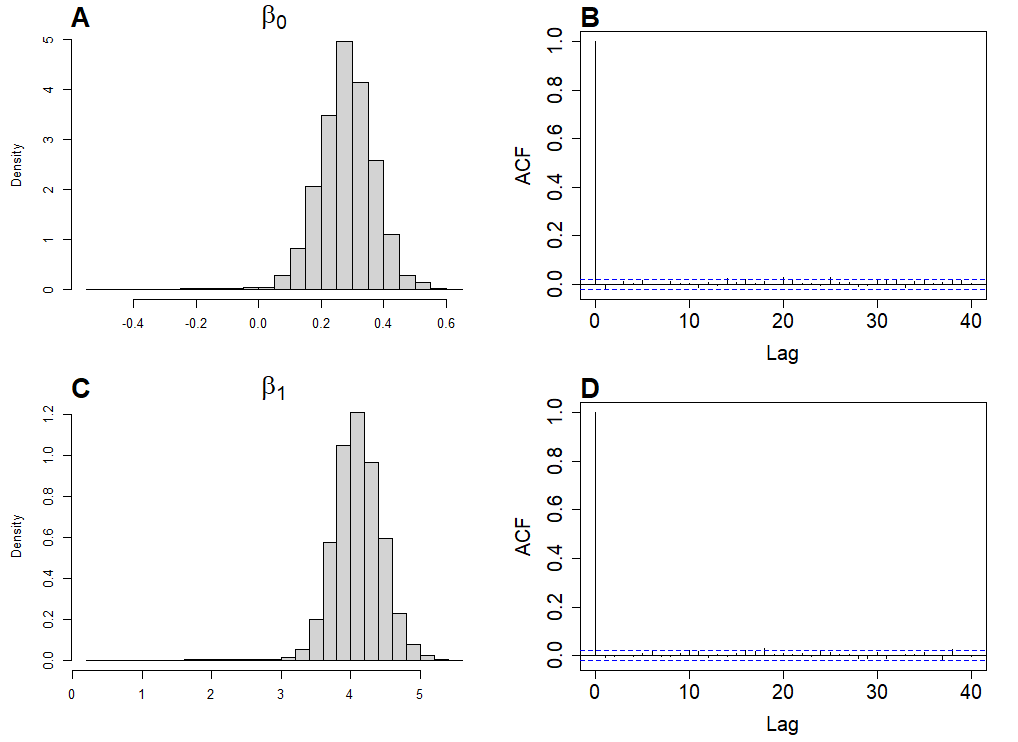}}
\caption{\textbf{A}-\textbf{D} Histograms and ACFs of the posterior distributions
of the FBR parameters $\beta_0$ and $\beta_1$.}
\label{fig_beta1_beta2}
\end{figure}

As a burning strategy we discarded the first 4000 samples of each chain.
The remaining samples passed through the following thinning procedure to
ensure no serial dependence between the posterior samples. The empirical 
autocorrelation function (ACF) applied to each chain revealed different dependence
levels (significant lags); we found 15 as the largest dependence level.
Our thinning strategy was to segment the sequence of samples in non overlapping
intervals such that 
the last member of a given interval
and the first element of the next one were separated at least 15 units. 
Then, at random we selected some samples from these strategically separated
intervals; we took 500 non correlated samples on each chain. 
Thus, for each parameter of the FBR model we ended up producing 10000 
posterior samples. Figure~\ref{fig_p_phi} depicts posterior samples
of $p$ and $\phi$ along with their corresponding ACFs. 
Similarly, Figure~\ref{fig_beta1_beta2} portraits the histogram of the posterior
distribution of the flexible Beta regression parameters $\beta_0$ and 
$\beta_1$ accompanied by their ACFs. 

%
We consider that our implementation of the nested Metropolis-Hastings
and Gibbs sampling algorithm and the subsequent application of our thinning approach
produce appropriate posterior samples of the FBR model.
In order to fit the FBR model we proceeded as explained at the end of 
Section~\ref{ss_MH_Gibbs}. 
%

\begin{figure}[htbp]
\centerline{\includegraphics[width=0.5\textwidth]{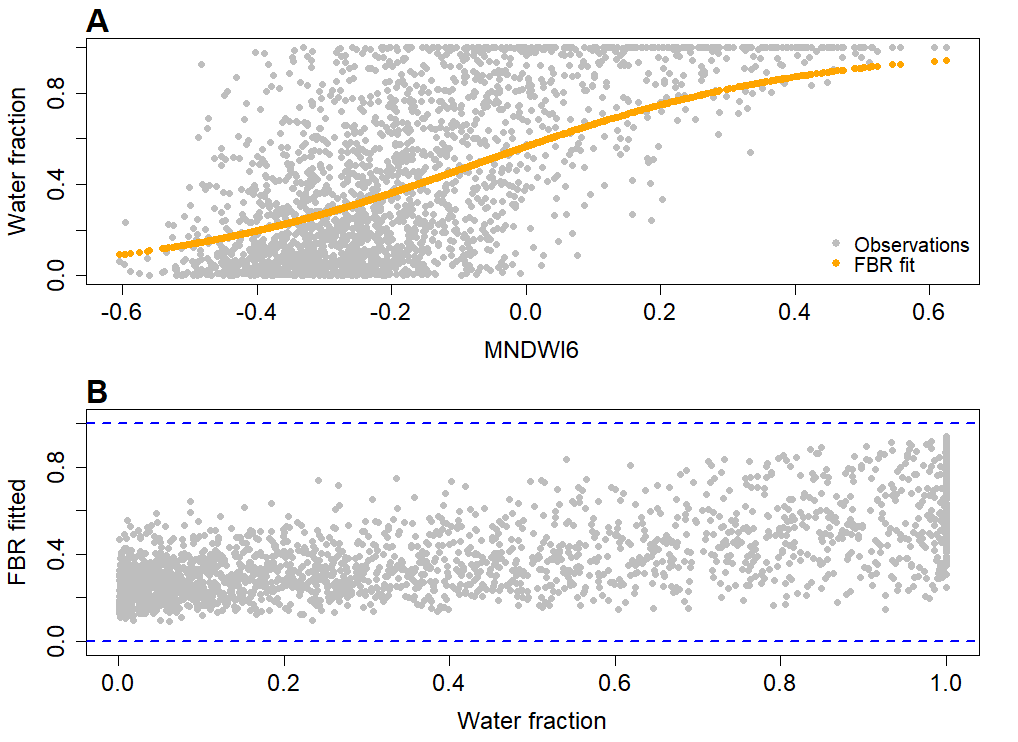}}
\caption{\textbf{A} Regression function $\hat{\lambda}_2$ (in orange) fitted via the FBR model. \textbf{B} Scatterplot of water fraction vs. FBR fitted values.}
\label{fig_FBR}
\end{figure}

Figure~\ref{fig_FBR}-\textbf{A} shows the $\hat{\lambda}_2$ regression function
fitted by the FBR model. As it is shown in Figure~\ref{fig_FBR}-\textbf{B}
the predicted values of this model lie on the interval (0,1), satisfying
the minimal requirement of a predictor of water fraction. 
The FBR model presented in this work is suitable for independent observations,
which is a strong assumption to hold true on satellite-derived variables.
In particular, the scatterplot of Figure~\ref{fig_FBR}-\textbf{A} shows
levels of heteroscedasticity in the measurements.
As argued by \cite{migliorati2018} the FBR can be used on heteroscedastic
observations by allowing some parameters, for instance the precision 
parameter $\phi$, to depend on covariates. This modeling approach requires
further study and it will be explored in the future.



\section{Conclusions}\label{sec.conclusion}

Modeling water fraction measurements derived from satellite images 
can be construed as the first step in measuring
the dynamic of open water bodies. In this work we presented
the flexible Beta regression model as an alternative to predict water fraction
as a function of the water index MNDWI6. 

We have provided an algorithm based on Bayesian estimation principles to
fit the FBR model. We consider that our implementation yields appropriate
results and is available to any interested user. 
We have shown that the predicted values of the FBR model lie on the interval $(0,1)$.
In this regard, this model improves the prediction made by the linear model.

Arguably, measurements of water fraction and water index are both spatially
related. It is sensible to expect that this relation is still present when
the measurements are vectorized (which is the type of data that we employed 
in this work). Our approach lacks of accounting for this apparent 
dependence between observations. 
It has been noticed by 
\cite{migliorati2018} that by allowing the precision parameter to depend
on covariates it is plausible that the FBR model may be able to account for
heteroscedasticity in the observations. We envision to explore
this idea in our future research. 

Subsequent steps for modeling open water fraction include
the acquisition of 
biophysical and geographical variables such as precipitation, temperature, 
elevation and hydrological basin models and include them into our model.
We believe that with this extra information along with the improvements
discussed above our model will become a useful tool to assess
the dynamics of the open water bodies of Mexico.

\section{Acknowledgment}

We are grateful to Berenice V\'azquez with CONABIO for kindly
providing the images analyzed in this paper. Julia Trinidad Reyes was
partly funded by Academia Mexicana de Ciencias through a summer grant
during ``Verano de la Investigaci\'on Cient\'ifica 2020''.


\ifCLASSOPTIONcaptionsoff
  \newpage
\fi

\end{document}